# Label Consistent Transform Learning for Hyperspectral Image Classification

Jyoti Maggu, Hemant K. Aggarwal, and Angshul Majumdar, *Senior Member, IEEE*

*Abstract*—This work proposes a new image analysis tool called Label Consistent Transform Learning (LCTL). Transform learning is a recent unsupervised representation learning approach; we add supervision by incorporating a label consistency constraint. The proposed technique is especially suited for hyper-spectral image classification problems owing to its ability to learn from fewer samples. We have compared our proposed method on state-of-the-art techniques like label consistent KSVD, Stacked Autoencoder, Deep Belief Network and Convolutional Neural Network. Our method yields considerably better results (more than 0.1 improvement in Kappa coefficient) than all the aforesaid techniques.

*Index Terms*—Transform learning, Dictionary learning, Classification

## I. INTRODUCTION

REPRESENTATION learning has seen a plethora of applications in hyperspectral image classification in the past few years. A large number of papers have used dictionary learning based approaches. There are several studies that applied deep learning techniques as well.

The precursor of such representation learning based techniques was the sparse representation based classification approach [1]. This has been heavily adopted by the remote sensing community [2-5]; mostly owing to its non-parametric nature. We have cited only a few studies, owing to the space constraints; there are a large number of other papers on this topic.

In sparse representation based classification, the raw training samples act as basis for each class. Dictionary learning based techniques replace the raw samples by a learnt dictionary; many papers in hyperspectral image classification are based on this basic premise [6-9].

Label consistent KSVD [10] proposed a dictionary learning technique that had an in-built classifier. Since its inception, it has been used in various domains for simultaneous representation learning and classification. There are several papers on hyperspectral image classification based on this approach [11-13].

Deep learning based techniques are also gaining popularity

J. Maggu is with Indraprastha Institute of Information Technology, Delhi, India (e-mail: jyotim@iiitd.ac.in).
H. K. Aggarwal is with Indraprastha Institute of Information Technology, Delhi, India (e-mail: hemanta@iiitd.ac.in).
A. Majumdar is with Indraprastha Institute of Information Technology, Delhi, India (e-mail: angshul@iiitd.ac.in).

in the remote sensing community. In [14] stacked autoencoder (SAE) with logistic regression has been used. Deep belief network (DBN) based hyperspectral image classification has been carried out in [15]. In [16, 17] convolutional neural network (CNN) has been used for the same problem.

Our reference list on dictionary learning and deep learning based techniques in hyperspectral image classification is far from encyclopedic. We mention some of the major works in their respective areas. In a recent work, a new approach that combines deep learning with dictionary learning has been proposed [18]. It has been used for hyperspectral image classification [19].

Transform learning [20-22] is a new representation learning technique; it is the analysis version of dictionary learning. So far it has mainly been used for solving inverse problems [20, 21, 23]. A handful of short papers used it for feature extraction in a naïve fashion [24-26]. In this work, for the first time we propose a supervised version of transform learning. This is based on the label consistency criterion introduced in [10].

We will show that label-consistent transform learning yields significantly superior results compared to dictionary learning and deep learning based methods. Transform learning is less prone to over-fitting compared to other approaches. Since the amount of training data is always a constraint in hyper-spectral imaging, our method excels over the rest in this scenario.

## II. LITERATURE REVIEW

### A. Label Consistent Dictionary Learning

Dictionary learning is a synthesis approach; it learns a dictionary ($D$) from the data ($X$) such that it can synthesize / generate the data from the leant coefficients ($Z$). Mathematically, this is expressed as,

$$X = DZ \qquad (1)$$

This is a matrix factorization problem. In early days, it was solved using alternating minimization. There were no constraints on the dictionary atoms or the coefficients. In recent times, 'sparse coding' enforces a sparsity constraint on the coefficients. K-SVD is the most popular approach to solve the ensuing problem. This is expressed as,

$$\min_{D,Z} \|X - DZ\|_F^2 \text{ such that } \|Z\|_0 \le \tau \qquad (2)$$

The $l_0$-norm promotes sparsity in the coefficients; the parameter $\tau$, controls the level of sparsity.

Sparse coding finds a plethora of applications in image processing, especially in the solution of inverse problems. There is also a large volume of work on supervised dictionary learning based methods; the main idea in such studies is to enforce class-wise discrimination into the coefficients Z.

Discriminative [27] / Label-consistent dictionary learning [10] is one of the most popular approaches (the equivalence between the two has been proven in [28]) in supervised dictionary learning. Not only does it learn the features in a supervised fashion, it also learns a linear classifier. The mathematical expression is given by,

$$\min_{D,Z,M} \|X - DZ\|_F^2 + \lambda \|Q - MZ\|_F^2 \text{ such that } \|Z\|_0 \leq \tau \quad (3)$$

Here $M$ is the linear mapping that is learnt between the coefficients $Z$ and the binary class labels $Q$.

During training, this optimization problem (3) is solved. For testing, the feature for the test sample $x_{test}$ is generated by sparse coding.

$$\min_{z_{test}} \|x_{test} - Dz_{test}\|_2^2 \text{ such that } \|z_{test}\|_0 \leq \tau \quad (4)$$

The generated features is multiplied by the linear map to produce the corresponding target. However the target is usually not binary, hence the position of the maximum value is taken as the class of the test sample.

### B. Transform Learning

Transform learning is relatively recent. Hence, we discuss it in detail. Transform learning analyses the data by learning a transform / basis to produce coefficients. Mathematically this is expressed as,

$$TX = Z \quad (5)$$

Here $T$ is the transform, $X$ is the data and $Z$ the corresponding coefficients. The following transform learning formulation (6) was proposed in [20, 21] –

$$\min_{T,Z} \|TX - Z\|_F^2 + \lambda \left( \|T\|_F^2 - \log \det T \right) + \mu \|Z\|_1 \quad (6)$$

Here the parameters ($\lambda$ and $\mu$) are positive. The factor $-\log \det T$ imposes a full rank on the learned transform; this prevents the degenerate solution ($T=0, Z=0$). The additional penalty $\|T\|_F^2$ is to balance scale; without this $-\log \det T$ can keep on increasing producing degenerate results in the other extreme.

In [20, 21], an alternating minimization approach was proposed to solve the transform learning problem. This is given by –

$$Z \leftarrow \min_Z \|TX - Z\|_F^2 + \mu \|Z\|_1 \quad (7a)$$

$$T \leftarrow \min_T \|TX - Z\|_F^2 + \lambda \left( \|T\|_F^2 - \log \det T \right) \quad (7b)$$

Updating the coefficients (7a) is straightforward. It can be updated via one step of soft thresholding. This is expressed as,

$$Z \leftarrow signum(TX) \cdot \max \left( 0, abs(TX) - \mu \right) \quad (8)$$

Here '$\cdot$' indicates element-wise product.

In the initial paper on transform learning [20], a non-linear conjugate gradient based technique was proposed to solve the transform update. In the more refined version [21], with some linear algebraic tricks they were able to show that a closed form update exists for the transform.

$$XX^T + \lambda I = LL^T \quad (9a)$$

$$L^{-1}XZ^T = USV^T \quad (9b)$$

$$T = 0.5R\left(S + (S^2 + 2\lambda I)^{1/2}\right)Q^T L^{-1} \quad (9c)$$

The first step is to compute the Cholesky decomposition; the decomposition exists since $XX^T + \lambda \varepsilon I$ is symmetric positive definite. The next step is to compute the full SVD. The final step is the update step. The proof for convergence of such an update algorithm can be found in [22].

There are only a handful of papers on this topic. Theoretical aspects of transform learning are discussed in [20-22]. In [23] it is used to solve inverse problems. Exactly the same formulation has been dubbed as 'analysis sparse coding' when applied to feature generation [24].

### III. LABEL CONSISTENT TRANSFORM LEARNING

Today dictionary learning is a popular representation learning tool. A short analysis shows that for a synthesis dictionary of size $m \times n$, with sparsity (number of non-zero elements in Z) $k$, the number of sub-spaces is $^nC_k$ for $k$-dimensional sub-spaces. For analysis transform learning of size $p \times d$, with co-sparsity $l$ the number of sub-spaces is $^pC_l$ for sub-spaces of dimension $d$-$l$. If we assume equal redundancy, i.e. $p=n=2d$, and equal dimensionality of the sub-space, i.e. $k=d$-$l$, the number of analysis sub-spaces will be $n$ where as the number of synthesis sub-spaces are $k \log_2(n/k)$ (via Stirling's approximation); usually $n \gg n/k)$. For example with $n=700$, $l=300$ and $k = 50$, the number analysis sub-spaces are 700 whereas the number of synthesis sub-spaces are only 191.

The aforesaid discussion means that for a transform and a dictionary of same dimensions, an analysis transform is able to capture significantly more variability in the data compared to a synthesis dictionary. In other words, for a fixed training set a smaller sized transform need to be learned compared to a dictionary. From the machine learning perspective, given the limited training data, learning fewer parameters for the transform has less chance of over-fitting than learning a larger number of synthesis dictionary atoms. Hence, for limited training data, as is the case in hyper-spectral image classification, transform learning can be assumed to yield better generalizability (and hence better results) compared to dictionary learning. Hence, we propose to base our work on transform learning.

In label consistent transform learning, the transform operates on the data to generate the coefficients ($TX=Z$), one also learns a linear classifier that maps the learnt coefficients to the binary class labels: $Q=MZ$. Combining the two terms – data fidelity $\|TX - Z\|_F^2$ and label consistency $\|Q - MZ\|_F^2$ along with the associated penalties (on transform and coefficients) we arrive at the following formulation:

$$\min_{T,Z,M} \|TX - Z\|_F^2 + \|Q - MZ\|_F^2$$
$$+ \lambda \left( \|T\|_F^2 - \log \det T \right) + \mu \|Z\|_1 \quad (10)$$

We employ the alternating direction method of multipliers

(ADMM) [29] approach to segregate (10) into the following (easier) sub-problems.

$$T \leftarrow \min_{T} \|TX - Z\|_F^2 + \lambda \left( \|T\|_F^2 - \log \det T \right) \quad (11a)$$

$$Z \leftarrow \min_{Z} \|TX - Z\|_F^2 + \|Q - MZ\|_F^2 + \mu \|Z\|_1$$

$$\equiv \min_{Z} \left\| \begin{pmatrix} TX \\ Q \end{pmatrix} - \begin{pmatrix} I \\ M \end{pmatrix} Z \right\|_F^2 + \mu \|Z\|_1 \quad (11b)$$

$$M \leftarrow \min_{M} \|Q - MZ\|_F^2 \quad (11c)$$

The update for the transform remains exactly the same as before (9). The update for the linear map $M$, remains the same as in coupled dictionary learning. It can be obtained via the pseudo-inverse. Update for $Z$ is slightly different from the usual transform learning; but nevertheless it is an $l_1$-regularized least squares problem and hence can be updated via iterative soft-thresholding [30].

This concludes the training stage. During testing, given the test sample $x_{test}$, we first need to generate the corresponding coefficients using the learned transform. This is given by,

$$\min_{z_{test}} \|Tx_{test} - z_{test}\|_F^2 + \mu \|z_{test}\|_1 \quad (12)$$

This has a closed form update –

$$z_{test} \leftarrow signum(Tx_{test}) \cdot \max\left(0, abs(Tx_{test}) - \mu\right) \quad (13)$$

Notice that, owing to the closed form update of the transform coefficients during testing, generating features is much faster than dictionary learning. Dictionary learning requires solving an expensive iterative optimization problem ($l_1$-minimization) during testing.

Once the coefficient vector is obtained, it is multiplied by the linear map to produce the approximate class label:

$$\hat{q} = Mz_{test} \quad (14)$$

Obviously the obtained label is not a binary vector. But the class of the test sample can be identified by finding the index of the maximum coefficient in $\hat{q}$.

In general label consistent transform learning has a much faster operation than its dictionary learning version. The testing stage requires two matrix vector multiplications only. In dictionary learning, one requires solving an iterative optimization problem. Thus label consistent transform learning can be used for real-time analysis applications.

## IV. EXPERIMENTAL EVALUATION

We evaluate our proposed technique on the problem of hyperspectral image classification; the datasets are Indian Pines which has 200 spectral reflectance bands after removing bands covering the region of water absorption and 145*145 pixels of sixteen categories, and the Pavia University scene which has 103 bands of 340*610 pixels of nine categories. In this work we follow the standard evaluation protocol on these datasets. For both the first datasets, we randomly select 10% and 2% of the labelled data as training set and rest as testing set for the Indian Pines and Pavia respectively; this is the standard evaluation protocol.

In this work we have compared with several state-of-the-art techniques – LC-KSVD [11], stacked autoencoder (SAE) with logistic regression [14], deep belief network (DBN) with logistic regression [15] and convolutional neural network (CNN) [16]. The configuration of the methods compared against are from the corresponding studies. We do not discuss them owing to limitations of space.

For our proposed method we have used a transform with 40 basis elements. The values of the parameters used are λ=0.1 and μ=0.05. These values have shown to yield good results for all the datasets. We did not fine tune the parameters to yield the best results in each.

The input features used in this work are raw pixel values – this is the input for all the techniques. We do not do any pre-processing or post-processing. The detailed experimental results are shown in Tables I and II.

TABLE I
CLASSIFICATION RESULTS ON INDIAN PINES

| Class | # Training | # Testing | Total | Proposed | LC-KSVD | SAE | DBN | CNN |
|---|---|---|---|---|---|---|---|---|
| 1 | 15 | 31 | 46 | 83.87 | 64.52 | 22.58 | 24.10 | 66.31 |
| 2 | 142 | 1286 | 1428 | 79.47 | 39.50 | 60.73 | 62.21 | 60.26 |
| 3 | 83 | 747 | 830 | 65.33 | 23.83 | 27.84 | 27.84 | 30.49 |
| 4 | 50 | 187 | 237 | 77.01 | 23.53 | 25.13 | 26.46 | 31.09 |
| 5 | 48 | 435 | 483 | 88.05 | 53.79 | 80.92 | 82.40 | 81.47 |
| 6 | 73 | 657 | 730 | 98.63 | 67.58 | 92.24 | 93.74 | 95.11 |
| 7 | 20 | 8 | 28 | 100.00 | 37.50 | 75.00 | 75.00 | 100.00 |
| 8 | 47 | 431 | 478 | 94.20 | 84.69 | 98.14 | 96.62 | 94.20 |
| 9 | 15 | 5 | 20 | 100.00 | 60.00 | 60.00 | 60.00 | 80.00 |
| 10 | 97 | 875 | 972 | 69.49 | 30.29 | 38.86 | 40.38 | 36.66 |
| 11 | 245 | 2210 | 2455 | 78.60 | 69.41 | 82.13 | 83.64 | 80.49 |
| 12 | 59 | 534 | 593 | 68.73 | 26.03 | 41.57 | 43.10 | 51.03 |
| 13 | 20 | 185 | 205 | 96.76 | 36.22 | 95.68 | 96.15 | 96.49 |
| 14 | 126 | 1139 | 1265 | 91.75 | 92.45 | 95.26 | 93.77 | 92.13 |
| 15 | 38 | 348 | 386 | 50.00 | 34.77 | 30.46 | 31.95 | 40.10 |
| 16 | 50 | 43 | 93 | 97.67 | 93.02 | 97.67 | 97.67 | 93.16 |
| OA | | | | 79.84 | 55.02 | 68.19 | 69.68 | 82.88 |
| AA | | | | 83.72 | 52.32 | 64.01 | 66.11 | 70.56 |
| Kappa | | | | 0.77 | 0.48 | 0.63 | 0.66 | 0.67 |

TABLE II
CLASSIFICATION RESULTS ON PAVIA

| Class | # Training | # Testing | Total | Proposed | LC-KSVD | SAE | DBN | CNN |
|---|---|---|---|---|---|---|---|---|
| 1 | 132 | 6499 | 6631 | 95.56 | 64.86 | 89.21 | 88.76 | 90.91 |
| 2 | 372 | 18277 | 18649 | 97.20 | 81.93 | 98.48 | 97.29 | 93.42 |
| 3 | 41 | 2058 | 2099 | 81.75 | 60.48 | 24.34 | 25.86 | 58.51 |
| 4 | 61 | 3003 | 3064 | 94.96 | 75.16 | 74.16 | 75.40 | 75.16 |
| 5 | 26 | 1319 | 1345 | 99.50 | 99.83 | 98.86 | 98.29 | 98.88 |
| 6 | 100 | 4929 | 5029 | 87.25 | 55.07 | 22.66 | 23.64 | 54.03 |
| 7 | 26 | 1304 | 1330 | 81.20 | 65.08 | 30.31 | 32.61 | 60.72 |
| 8 | 73 | 3609 | 3682 | 81.23 | 74.32 | 58.69 | 60.45 | 68.06 |
| 9 | 18 | 929 | 947 | 100.00 | 84.06 | 60.18 | 61.26 | 81.19 |
| OA | | | | 93.12 | 74.02 | 63.13 | 65.03 | 76.98 |
| AA | | | | 90.96 | 73.42 | 61.87 | 62.62 | 75.65 |
| Kappa | | | | 0.91 | 0.66 | 0.61 | 0.62 | 0.76 |

One must note that the results shown here cannot be directly compared with those in [11, 14-16]. This is because we do not do any post-processing or pre-processing. In [11], as a post processing step, the classification results are pooled using nearest neighbor in order to enforce local consistency. In [14, 15], raw pixel values are not used as input features. They use spatio-spectral features, which is a PCA applied on the pixel values across all spectral bands. Besides, the train and the test sets used in [14-16] are different from the ones used here. They assumed 80%-90% labeled data and only tested on the remaining 10%-20% samples. This is not a typical scenario in hyper-spectral image classification; usual protocols do not use so much labeled training data (e.g. [5]).

Our evaluation protocol is uniform and tests the raw analysis capability of the different methods. We find that our method yields results that are significantly superior compared to the state-of-the-art techniques in all possible measures (overall accuracy, average accuracy and Kappa coefficient). This is because all the techniques compared against (LC-KSVD, SAE, DBN, CNN) are data hungry. On limited training data they over-fit. LCTL on the other hand does not; and hence yields very good results.

For visual evaluation, we show the classification results from different techniques in Fig. 1; this corresponds to the Pavia dataset. The images corroborate the numerical results. Owing to limitations in space we do not show the results for Indian Pines. But the conclusions drawn therein are similar.

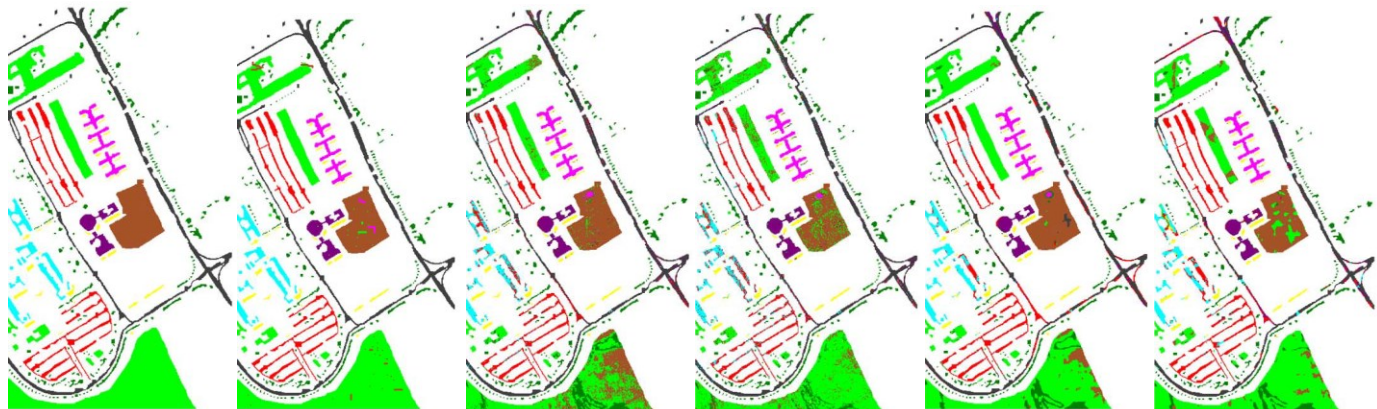

Fig. 1. Left to Right – Groundtruth, Proposed, LC-KSVD, SAE, DBN and CNN

Finally we show the training and testing times of our proposed approach. We compare it with LC-KSVD. The times for deep learning methods are significantly larger. We show the runtime for DBN since it is the fastest of all the deep learning techniques. The results are shown in the following table. All the experiments have been carried out on a Windows 10 PC running on an intel i7 with 16 GB RAM. The experiments were conducted on Matlab 2012a.

TABLE III
COMPARISON OF RUNTIMES IN SECONDS

| | | LC-KSVD | Proposed | DBN |
|---|---|---|---|---|
| Indian Pines | Testing | 2.75 | 0.01 | 0.52 |
| | Training | 34.87 | 0.21 | 523.56 |
| Pavia | Testing | 19.01 | 0.18 | 0.76 |
| | Training | 255.73 | 0.56 | 1252.47 |

Our method is the fastest both in training and testing speed. During training, both LC-KSVD and our proposed method requires computing SVDs. But the dictionary learning based method is slower owing to two reasons. First, it takes more iterations to converge. Second, the required size of the dictionary is much larger than the size of the transform. This has been discussed before. The analysis transform can capture more variability in a smaller size, compared to a dictionary. During testing, the dictionary learning based method needs solving an iterative optimization problem; for us it is only a matrix vector product. Hence our method is two orders of magnitude faster.

The DBN has a significantly slower training time. But it has a faster testing time compared to dictionary learning; this is because they only need a few matrix vector multiplications. Owing to the deeper architecture, the number of matrix products DBN needs to compute is more than ours, hence it is slower than our proposed technique.

V. CONCLUSION

This work proposes a new tool for image classification. It is based on the transform learning formulation. We proposed to supervise transform learning by adding a label consistency constraint – an idea that has enjoyed significant success in recent years on the dictionary learning framework. Our formulation is called label consistent transform learning (LCTL).

There are two advantages of LCTL over its dictionary learning counterpart [10, 27, 28]. The first one is theoretical. Transform learning can learn from far fewer samples compared to dictionary learning; i.e. for fewer training samples transform learning will not over-fit but dictionary learning will. This makes it especially suitable for hyperspectral image classification problems, since the number of labeled training samples are always limited.

The second advantage is computational speed. In the testing stage, transform learning has a closed form solution for generating coefficients. And for our formulation, the analysis requires a simple matrix vector multiplication. Dictionary learning on the other hand requires solving an iterative optimization problem. Thus transform learning can operate in real-time whereas dictionary learning cannot. Although time is not of essence in hyperspectral image classification problems, but may be useful in other areas – such as tracking.